# ANALYSIS AND KNOWLEDGE DISCOVERY FROM SENSORS DATA TO IMPROVE ENERGY EFFICIENCY

Xavier Vasques, PhD / IBM IT Specialist, Xavier.Vasques@fr.ibm.com
Thibaut Possompes / PhD Student, thibaut.possompes@fr.ibm.com
Herve Rey / IBM IT Architect
Marine Le Touzé, ENSIMAG, IT Specialist
Nicolas Auboin, IBM IT Specialist
Emmanuelle Passot, IBM Specialist
Benoit Lange / PhD Student
*IBM Montpellier, Product and Solution Support Center (PSSC), Innovation Lab, Montpellier, France*

**ABSTRACT**

Increases in energy prices and the global goal of mitigating $CO_2$ emissions necessitate the development of intelligent Building Management Systems (BMS) that operate on an energy-efficient basis. Data Centers, buildings and/or group of buildings are often responsible for huge energy consumption. One way to monitor and optimize energy consumption is to instrument buildings using sensors (temperature, pressure, humidity …) in order to track and solve wrong usage of energy management systems. The majority of the BMS are processing the data dynamically without taking into account the data history due to their constraint problems (time, bandwidth and calculation capability) and data resources. The RIDER project brings together a consortium of research laboratories and enterprises including IBM, to share their expertise in research and development of smart Information Technology (IT) energy platforms. In this context, we aim to improve energy efficiency of buildings or group of building (including data centers) using IT. One of the objectives is to identify valid, potentially useful, and ultimately understandable patterns in data for improving energy efficiency. We propose in this paper an approach of using an integrated platform able to interconnect instrumented buildings and sites, and to provide a high-level point of view for increasing our knowledge from sensors. The expected results are to estimate physical parameters that influence energy consumption based on data set history. Different correlation could be found between different variables, for example, indoor air quality and energy consumption. These results could be applied at a location where no sensor is placed and predict energy consumption from different variables.

## 1. INTRODUCTION

Buildings are important contributors to total energy consumption in countries. Building energy demands can be reduced significantly due to developments in the field of mechanical and civil engineering. However, the conventional energy reduction solutions require consequent investment. In contrast, it is possible to achieve energy savings with minimal additional cost by giving more information to building management systems (BMS). Indeed, several parameters can be manage by the BMS including heating, ventilation, air conditioning systems (HVAC), blind positioning and lighting systems. Numerous articles exist on how to improve energy efficiency. One way to improve energy efficiency in building is to modify the occupant behavior by providing him electricity consumption feedback in order to provide motivation, guidance and verification. Berges et al. (Berges et al. 2008; Berges et al. 2010) use non-intrusive load monitoring (NILM) that is capable of gathering detailed energy-use data with minimal equipment cost and installation time. However, as explained by the authors, variations in measurements between metering devices complicate the process of compiling the necessary appliance profiles. The RIDER project, aims to show that the use of Information Technology and Communication (ICT) will permit to optimise the energy efficiency of different types of buildings or group of buildings, including data centers. This project brings together a consortium of research laboratories and enterprises to share their expertise in research and development of smart energy platforms. The goal is to develop an intermediary layer of optimisation, located between the existing mechanism at a building and those of the power distribution management networks. This intermediary layer will allow the management of different energy types, integrating conventional energy, renewable energy (thermal, photovoltaic or wind, etc.) but also internal energy generation such as the heat generated by a data center or an industrial process. The RIDER system aims at providing an integrated platform able to interconnect instrumented buildings and sites, and to provide a high-level point of view for increasing our knowledge on energy optimisation. The RIDER system is made exclusively of IT components to allow the capture and processing of real-time information necessary to control and optimise the entire system under consideration. The sensors and actuators are connected and managed by the BMS. The BMS are connected to RIDER. External information sources can also be connected to the system in order to extend the scope of RIDER data analysis. Information sources include, among others, weather forecast, room occupancy planning or social events planning. BMS gather data from the sensors placed in the building and send orders to the actuators. RIDER is able to perform data analysis and run algorithms in order to control more efficiently buildings. Not all BMS have the same functions, but they can easily implement several features including communication protocols, built-in web servers or data base access in order to be able to communicate with RIDER. RIDER is positioned to develop new domain specific methods of data processing, building modelling variants, algorithms, data visualisation techniques, communication protocols and associated architecture. RIDER will be used to implement a pilot, to demonstrate the viability of its design and implementation, by testing it initially in the Green Data Center of IBM Montpellier (France), Capdeville in Montpellier equipped with solar panels, and Archipel theater, a modern cultural building of large capacity in Perpignan (France).

In this paper, we propose an approach to federate the various components of instrumented buildings in order to process and analyse all the data coming from buildings in real time for energy optimisation.

## 2. CONTEXT PRESENTATION

RIDER, recognized by the DERBI French Cluster, is supervised by a consortium, led by IBM, of large enterprises and small & medium enterprises (EDF, Cofely, Pyrescom, Enoleo, Coronis, Asa) and university laboratories (University of Montpellier 2: LIRMM, IES; University of Perpignan: PROMES) in the Languedoc Roussillon (France) area. The goal of RIDER is to improve Energy Consumption through the use of IT platforms enabling the interaction between all energy consumers and providers, with the objective to optimize energy usage of a group of physical entities combining predictive modeling, real data captured through sensors & actuators technology, and business intelligence. ICT enables capturing, in real-time thousands of parameters describing the energy state of a given environment. ICT makes it possible to analyze and compare models and take actions to enable optimal use of all resources available in this environment, and thus reduce the overall consumption of energy (renewable and non renewable).

The project's main objective is to develop an intelligent platform for managing energy consumption aspects of multi-scale and multi-standard building complex with optimization of energy usage as a goal. The system can be summarized into 2 levels, the local and the global.

The local level deals with the cases of "isolated" smart buildings (Figure 1). For the purpose of this project, a smart building has an added layer of intelligence through which elements of a building including temperature, electricity, ventilation, water, waste management, telecommunications, and physical security can now be integrated for better management and control. Smarter, more sustainable, buildings can quickly sense and respond at every system level possible. The RIDER will capture the data to be processed in order to manage and optimize the use of energy in the isolated buildings independently.

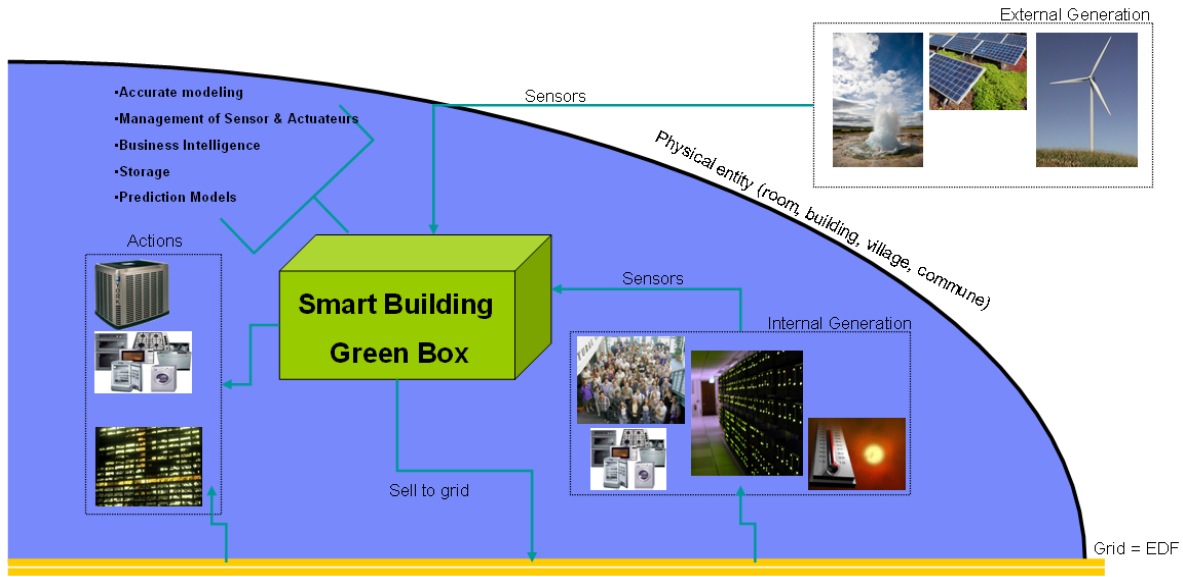

Figure 1: "isolated" smart building

Global level deals with the management and optimisation of energy usage globally in a federation of isolated smart buildings in an interconnected network (Figure 2).

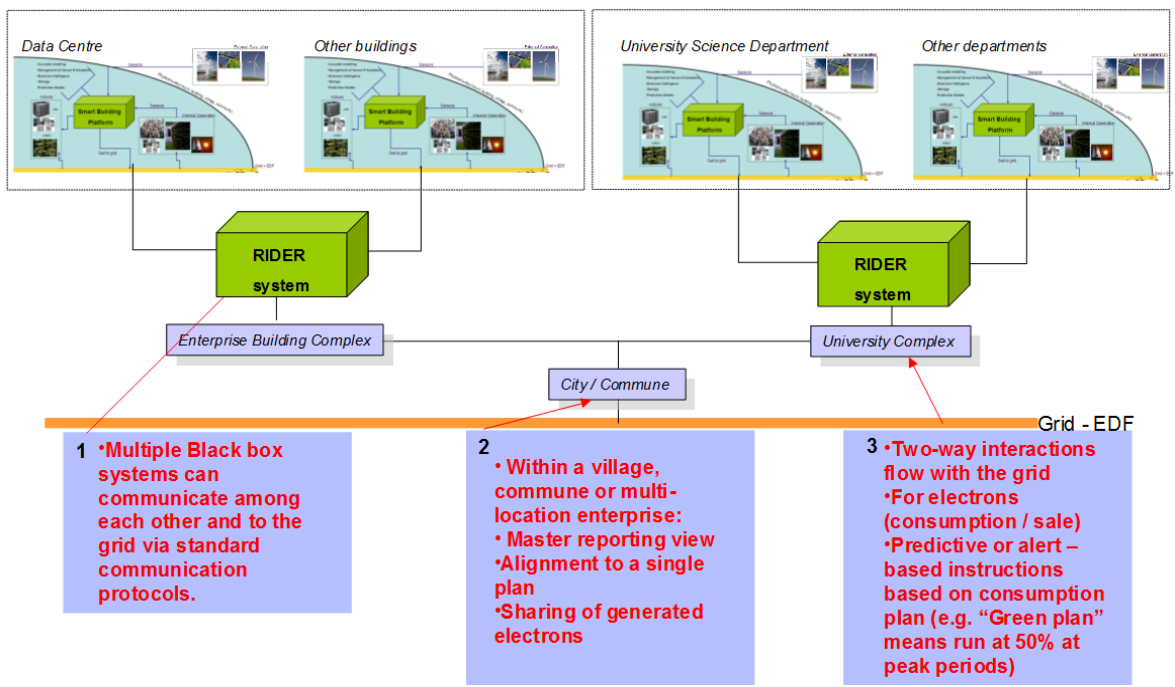

Figure 2: The interaction of managed buildings.

The BMS, connected to the RIDER system, manage sensors (internal and external information sources) and actuators. BMS gather all the data from the sensors placed in the building and send orders to the actuators (Figure 3). The BMS data structure is specific to each of them. Sensors and actuators can be referenced only by their memory address, or, for the most advanced BMS be linked to objects, e.g., HVAC systems, or a building blueprint. External information sources refer to any kind of service from which information can be gathered. At the building level, the RIDER system can manage different modelling standards, different building management systems, different kind of building (personal house, office building, data centre, etc.) or sites. At the neighbourhood level the client will instrument the environment of buildings and sites, and centralise specific data from the lower level RIDER systems in order to furnish higher level functions such as coordinating energy consumption among several houses or building.

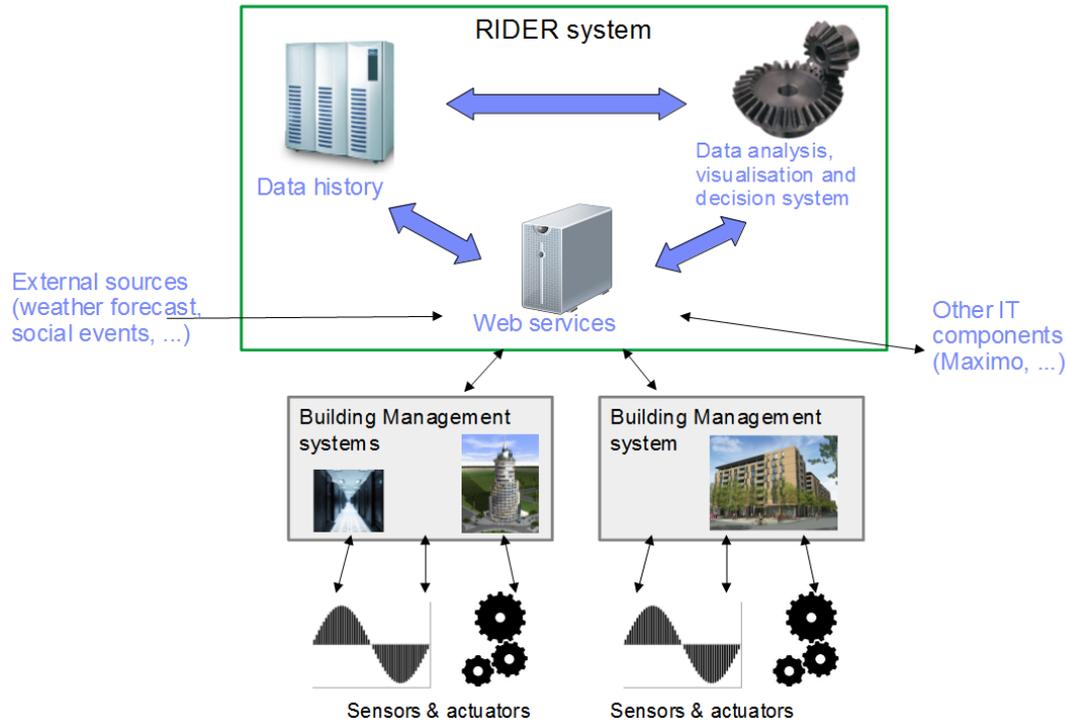

Figure 3. RIDER architecture overview – the system and its environment

The buildings are managed by one or several BMS communicating with the RIDER BMS interface which manage the data in real-time. An ad-hoc rule-based scenarios module is used to manage building data to perform energetic optimisation. These scenarios are based from notably data analysis including data mining, decision aid, statistical analysis (Cognos, SPSS), interactive data visualisation or stochastic differential equation (prediction). Additional modules can be loaded to perform computations on demand. The steps for creating scenarios are the following:
> 1. The scenarios must be described literally and identify, required infrastructure model objects, computations, input data, and output data.
> 2. The scenarios must be decomposed into rules and integrated in the scenario flow to make sure all objects are available.
> 3. Once the scenarios are described, the infrastructure model must be checked to ensure that all data can be found. If not, the model and those derived from it must be updated.
> 4. The scenarios can be written in JRules and integrated in the RIDER platform

For example, a scenario could be to determine when to start heating a room. The decision to heat is taken if there is someone in the room or if the presence planning of the room specifies that the room is occupied. The required data are: room temperature measure, room temperature set point, room presence schedule, room presence measure. This scenario is decomposed into 3 rules which are designed to work at different steps of the scenario flow. The step 1 occurs when new measures are loaded into the central memory. The step 2 is not required in this example. The step 3 consists in deciding if it is necessary to heat a room. Two rules evaluates whether it is necessary to heat accordingly to the presence sensor and the planning. If one of these rules determines that heating is required, an assertion is made to indicate that heating is required; otherwise the rule execution stops here. Step 4 determines and solves conflicting assertions, *e.g.,* it is necessary to heat and to chill. For the sake of simplicity, there is only one rule, which will check assertions and send a message if a room must be heated.

The implementation is currently being prototyped in a dedicated zone in industrial building containing offices, meeting rooms, a data center, and controllable HVAC systems.

The purpose of the RIDER project is to be used on multi-building infrastructures. To test all the features of such an infrastructure, the first RIDER Pilot uses the Green Data Center of IBM Montpellier, and a set of offices nearby. The Green Data Center and offices can be considered as two different buildings. Each of them is instrumented by a set of sensors permitting collection and analysis of data to determine the project issues.

A Green Data Center is a repository for storage, management, and dissemination of data for which the mechanical, lighting, electrical and computer systems are designed for maximum energy efficiency and minimum environmental impact. A data center is also producer of heat energy. The Green Data Center is managed and monitored (IT and non-IT) by using close to 200 probes and sensors installed throughout the center to gather real-time data. Several maps have been developed (thermal, pressure and hydrographic maps) that pro-actively detect any problems or deviations that would lower efficiency. For example, from the temperature sensors and statistical analysis, it is possible to predict the temperature evolution of the Data Center (by room, by rack, by servers ...). Over the day, we are able to predict the heat that we could recover from the racks for heating near offices and meeting rooms. We can anticipate and optimize the heating network usage (minimum usage of heater).

## 3. RIDER MODELLING

A scientific model is based upon empirical data which is acquired from various sensors, measures and observations. This data is structured thanks to a data model to facilitate its storage, access, and ensure its integrity. Sensor measures are gathered by BMS and sent to the RIDER platform, where it is structured accordingly to a data model. At this point, statistical analysis can be made on data in order to identify, for instance, correlations between sensor measures or observations (variables). Thereby a statistical model can be built with mathematical equations to describe how the observed building behave. Data mining algorithms can also be used to find recurrent patterns of data, and association rules between variables of the model. Similarly, physical modelling is based upon mathematical/physical equations describing a building behaviour accordingly to known thermal and physical properties. Both mathematical models resulting from statistics and physics can be seen as transfer functions representing the relation between input and output variables. Operational research can leverage the mentioned mathematical models in order to take optimal decisions accordingly to input variables and equations. This approach can be efficiently used to model one context such as an enterprise supply chain system, or a financial system of a customer, by refining the mathematical models to fit as well as possible to its characteristics. However, applying these models to another customer is a time consuming task because the mathematical models are bound to input and output data specific to one context. We have the same concern with mathematical models for buildings, but we want them to be easily applicable to other buildings. Furthermore, the mathematical models use different variables that designate very different concepts (material conductivity, historical data, data patterns, occupancy schedules, sensors measures, actuators commands, optimal set points to reach, …). Hence, a simple data model is not enough to share a common understanding of the domain among people. Each modelling field stay linked to the data it uses. We need to separate the domain knowledge from the operational knowledge to give freedom of expressiveness to the mathematical models and ease their interconnection. This is why a formal description of the manipulated concepts must be achieved and gathered in a knowledge base, currently called infrastructure model. A knowledge base is primarily used to enable making assumptions and inferring new information. We extend it by integrating processes and scenarios. Mathematical models can also be integrated in a knowledge base to help creating new information, which in turn enriches the knowledge base as a feedback loop, but also enable making high level scenarios. This knowledge base will help domain experts and users to easily understand the analysis, relationships, and decisions and to approve the results. 3D visualisation models will be able to represent not only data measures, but information about energetic efficiency and help the user to understand how the building can be optimised. The software model encompasses the descriptions of the required software to create a computer simulation based upon mathematical models, a knowledge base, and scenarios. It will be managed and organised with software product lines in order to customise it accordingly to customer requirements.

As depicted in figure 4 we developed an infrastructure model used to describe a building and its environment. It has been designed considering existing models such as the Industry Foundation Classes (IFC), but simplify them and add several concepts to focus on our platform needs. The infrastructure model aims at describing
– building blueprints, sensors, actuators
– indicators relevant in energetic optimisation scenarios, building analysis results, behaviour,

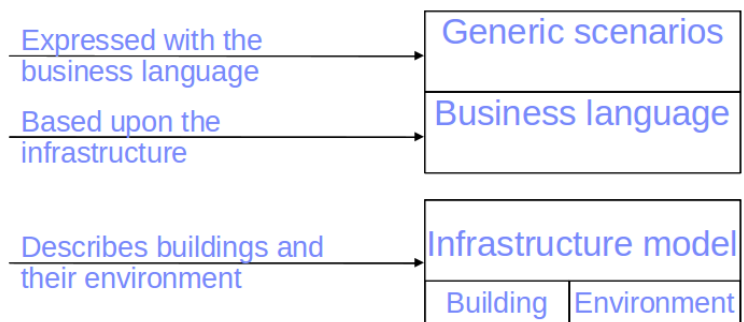

**Figure 4: Infrastructure model usage**

statistics. For example:
– the time to reach a temperature in a given room
– the optimal temperature set-point in an empty room
– savings achieved by RIDER
– statistical presence in a room
– virtual sensors
– multi-scale context, complexity abstraction concepts
– scale-specific indicators
– abstraction of lower-levels complexity
– HVAC control systems

This model is used as a pivot to exchange data between the various optimisation algorithms.

In the RIDER context, it's important to be able to achieve any comparison between various sensors values. A data warehouse is the best solution to do this type of analysis. The data warehouse equates with databases to make analysis easier. A metadata database is associated to the data warehouse to increase meaning of the data stored. Data are extracted from different supervisors. Each supervisor manages a kind of sensors. In data warehouse context, the ETL process extracts data from the staging area, transforms to standardise them and finally loads them into the data warehouse. Final step, the user can do easier and faster analysis, thanks to the data warehouse. To deduct a number of relevant analysis, a set of indicators has been identified. Theses indicators are used to make comparison in the analysis. Besides the values of sensors that can be considered indicators relevancy, a list of more advanced indicators was developed which can be compared to project requirements. The main goal of the Green Data Center pilot site is to implement a first version of RIDER system. The latter will be voluntary over instrumented in order to provide a prototype for the greatest number of possibilities tests.

This pilot will also help us to identify the real needs for implementing a RIDER system in any context. That is why a list of relevant indicators based on available attributes of databases has been created. This step was made in at the same time as the data warehouse design. Defining indicators permits us to be aware of the needs and the possibilities of the data warehouse. It was decided to gather all the sensors in a dimension in order to cross the maximum of the data. This generic modelling permits data crossing with indicators. By analogy, if we compare a dimension to an axis, the analysis of an indicator according to several dimensions is like a cube. For instance, it is possible to visualise the evolution of temperature by a sensor versus time in different areas. This analysis of the temperature is made on a time dimension and a location dimension. The data warehouse is constituted of the fact table which contains the measure, the acquired value. This will be the biggest table in the warehouse; it is the core of the warehouse. The latter gather references from dimensions. Specifics geographic and temporal dimensions are also used to localise and date a measure. Each measure is associated to a sensor and each sensor details all information about what it describes, for instance a machine (exceptions : relative humidity sensor for instance). In the data warehouse, sensors and actuators are merged. Indeed, in both cases, we will store values. For sensors, it will be measurement (e.g., temperature over time) and for the actuators, it will be their condition (e.g., the percentage of valve opening for air conditioning). Thus, to simplify the writing, the term "sensor" may refer a sensor or actuator. Virtual sensors values are computed from a set of other sensors. They are, like classical measures, located and dated. Thank to that, virtual sensors are modelled in the same way that conventional sensors. To have a data warehouse scheme as simple as possible, the type of architecture chosen is called "star scheme". This solution makes the analysis very simple because there is only one fact table and there is no hierarchy between the complex dimensions. To still be able to provide maximum information to those items (e.g., machines in the case of Green Data Center), all these information has been compiled in the items dimension connected to the concerned sensor in the fact table.

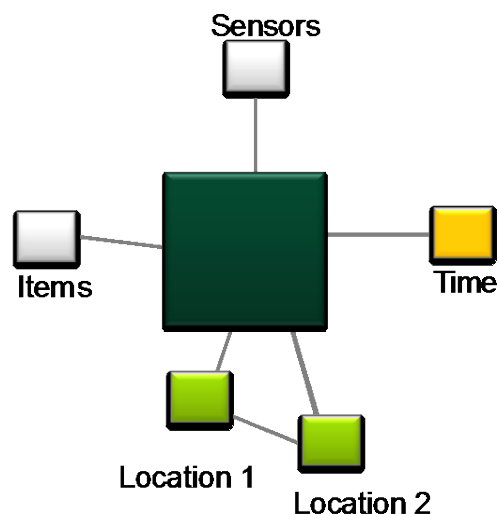

Figure 5: Simplified data warehouse scheme

# 4. RIDER WORK STREAMS

## 4.1. Software Product Lines

Here, we propose an approach to federate the various components involved in smart buildings, and to capitalise on the developed IT components by reusing them at a lower cost in the construction of future buildings. We achieve this goal by using software product lines methodology for:
- Gathering domain knowledge from the different stakeholders, and managing the whole software life-cycle of smart buildings related projects.
- modelling IT infrastructure necessary for instrumenting a building and analysing related data.

This leads us to describe how we could generate some parts of the IT infrastructure thanks to feature diagrams (Figure 6). We illustrate our approach in a model driven perspective and use the principles of the model transformation. Another key aspect of our project is to enable traceability through developed IT artefacts, domain models and feature diagrams in order to be able to manage domain specific and IT concepts as a whole.

| Domain engineering | Application engineering |
|---|---|
| **Application domain:** Elements to which features are associated with (*e.g.*, building optimisation feature is associated with the building class of the pivot model). | **Referenced objects:** When a feature has been chosen, this concept references to which objects the feature will have an impact on.<br><br>For instance, applying an energy optimisation scenario on one part of a building. |
| **Binding time:** Describes when the feature will have an impact, *e.g.*, compile time, server deployment time, scenarios execution, statistical model, ... | |
| **Validation:** Describes how the feature applicability will be controlled, *e.g.*, how to validate that a scenario will be applicable on a specific building. | |

Figure 6. Features can have the following concepts

Several people can be involved in the software product line creation. We can separate them in four categories: core IT platform development, product modules experts, business experts and final product customer.

## 4.2. Data Management

Here we focus on designing data warehouse architectural models for massive and heterogeneous data, to be integrated in a multi-building intelligent energy management platform. We rely on data from sensors, applications or external sources. Many stakes are involved in this project: high volume and importance of capitalising data, data seasonality, high data arrival rate akin to streams, heterogeneity of data (e.g. sensor data, data derived from uses, user preferences, etc..) monitored sites of various types with different energy parameters (e.g., Data center, residential space, production line) and finally strategic importance of making decisions in real time . It is necessary to get out of the positioning of the intelligence beyond the warehouse, and define a new approach that emphasises responsiveness of the system. To improve responsiveness of the RIDER system, it is necessary to "move the intelligence closer" to the data sources. This is why we will focus on the stage positioned between data sources and the warehouse: the ETL phase (Extract, Load and Load). The ETL is a process whose aim is to integrate data from operational databases in a warehouse and / or data marts. We investigate how to model and implement an ETL process which adapts to available data sources that may vary greatly, which can process "static" data as well as data flows, and which allows the implementation of real time reactivity. This process will thus embed stream mining tools in order to treat the data sources, especially to detect tendencies, exceptions, and drifts. We will focus more specifically within the ETL process on the mechanism of "data exception" or dismissed data, which involves extracting during the ETL process, data that does not respect the warehouse integration rules set a priori. These exceptions are now manually reprocessed and reintegrated into the cycle. The architecture we are working on will rely on the energetical data streams collected, that will be prepared and cleaned using ETL tools to be then loaded in a data warehouse. Before examining existing work on ETL, we first describe the work on data warehousing and data streams, before presenting the work dealing with temporality in data warehouses. We propose to treat them as a flow, classifying these exceptions by their nature, and initiating timely relevant action. Among these actions, analysis of deviant behaviors which would be detected as "incorrect" in a system whose rules are static, will allow us to detect the earliest changes in behaviors, and automatically update if necessary the corresponding rules to suit the energy behavior. This mechanism will apply to the data passing through the ETL in real time, but will be based on the results extracted from the warehouse using existing processes, including the collection of profiles to provide a positioning of the incoming data compared to the historical behavior of the system. The proposed architecture use collected energy data flows, formatted and integrated by the ETL process in a data warehousing structure.

### 4.3. Three dimensional visualisation of data

The goal is to produce a 3D visualisation for analysing and managing data coming from sensors providing different types of information (i.e. temperature, pressure, hygrometry …). In the 3D visualisation module, the sensors are placed in virtual rooms and the internal space is modelled using particles, as stated by the Industry Foundation Classes (IFC) model (a standard developed to allow interoperability in the building industry). The main constraint here is to produce a real time rendering. However, due to the number of vertices, latency appears. In our approach, we use a solution called LOD (Level Of Detail) to produce multi resolution 3D objects. J. Clark has introduced this solution in 1976. In this paper, J. Clark introduces the use of several mesh resolutions to simplify the 3D scene complexity. In our work, we use various simplification methods to provide interactive rendering. This solution allows rendering the most important part of data extracted from sensors.

Our approach produces an interpolation of sensors to obtain visualisation in 3D of temperatures (Figure 7). This approach produces an interpolation of temperatures in a realistic way. The usage of mathematical partitioning gives a first overview of the system. The first results show that Voronoï diagram is not enough in term of rendering. The only particles with a good weight are located between two sensors. This particles have correct distance weight. The other solution using a triangulation algorithm gives better results. It became easier to compute the location of the particle compared to the previous solution. The weight can be computed easily and if a particle is own by none of the sensors, Voronoï algorithm is used. Results are interesting to visualise influence of sensor in a room. The implementation of client server paradigm allows improving the rendering of the data. It consists of computing the expensive functions on a High Performance Computing system. It allows getting 24 FPS in the viewer. Before this step, 11 FPS were rendered. This improvement was useful and the developed communication protocol was low cost.

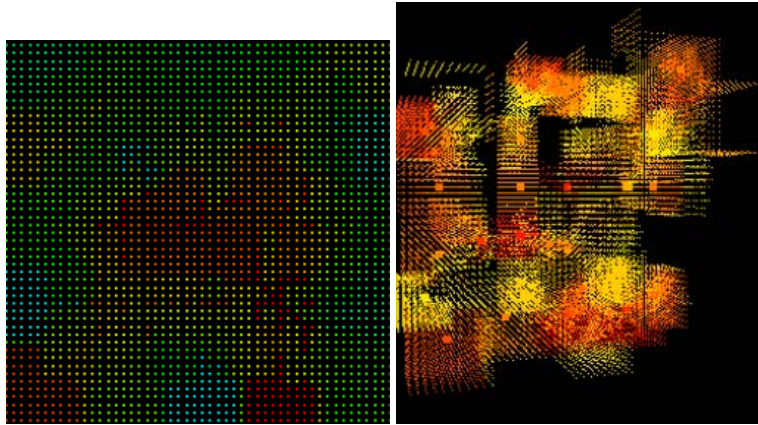

Figure 7. Our approach produces an interpolation of sensors to obtain visualisation in 3D of temperatures

**4.4. Micro energy sources and sensor's autonomy**

Recent development in electronics (e.g., ultra-low-power micro controller) and wireless communication has helped in development of sensor devices. Wireless sensor networks are now being widely used for different applications: dwelling, buildings, industry, transport, telecom, etc and the greatest problem faced with them is the lifetime of the sensor nodes as they may not be in a position to have wired power supply and they run on the onboard battery.
Batteries present another disadvantage; the increasing number of battery-powered portable products is creating an important environmental impact. The uses of micro energy sources to generate electricity have promising concept. Different Renewable energy that is being harvested to generate electricity includes energy domains within any typical environment, whether internal or external. Solar energy/Ambient Light, Temperature Gradients, Human Power, Air Flow, Pressure Gradients, Vibrations etc. are examples that we can envisage as possible sources for harvesting electrical energy from a typical environment.

The objectives of this study are the achievements of demonstrators and first characterizations of electrical converters of photovoltaic, thermoelectric, piezoelectric (PZE), pyrolectric (PRE), radio frequency type carried out using bulk materials and thin layers for their integration in microsystems or microsensors. With this intention our research group from its control of thin layers deposition techniques, from its knowledge of III-V materials, $V_2$-$VI_3$ (thermoelectric), piézo and pyrolectric and its aptitude to develop MEMS (cf. thermal accelerometer (3 patents), is in an ideal position to work out, study and characterize some of these direct converters of energy based on the physical principles mentioned above.

It will initially be necessary to define a list of sensors (catalogs) that each member of the consortium will be able to use in order to choose the sensor and the communication system best adapted with each environmental situation and of measurement.
After that and having carried out a state of the art of the various modes of energy transformations, the student will define the feasibility of each one of these converters (micro sources).
The lines which follow, detail the specific procedures under consideration for each type of converter:
PV broad band spectrum (cell tandem): this type of converter being based primarily on p-n junctions containing antimonides, the student will take part in the realization and the study of deposits of materials and components containing antimonides by MBE technique.
TE: study of the performances of conversion of a microthermogénératrice cell. Feasibility study of active materials of this type of converter by one of the techniques available in the IES will be realized.
PZE and PRE: based on our research group experience on piézo and pyrolectric (ZnO, Tantalate of Li and Li Niobate) materials growth and study, we'll contribute to the development of PZE and/or PRE type microsoucies.

## 4.5. RIDER Component Model

The component model describes the structure of a system in terms of its software components with their responsibilities, interfaces, relationships, and the way they collaborate to deliver the required functionality (Figure 8).

This diagram is composed of 4 distinct components:
1. Infrastructure Services: the purpose of this function is the building management system also called supervisor. This infrastructure is able to manage the building in term of collecting data from sensors or component (e.g. heating system or room temperature) and also uses actuators to control the building devices.
2. Supervisor / Business Management Systems (BMS): The supervisors in all the system infrastructures gather all data and communicate it to the system. Depending on the monitored system and on the supervisor type, different types of data and data structures will be gathered. In order to be normalised, the transmitted data needs to be processed later.
3. BMS: Physical infrastructures can be monitored.
4. Data Services:
    - Generic data Model: It defines the data model that must be used by all modules of the project. All data exchanges should be based upon it in order to facilitate the integration of new components. It could be used as a reference for XML formatting. The generic data model can be extended to take into account new types of data.
    - Physical equations: They can be applied to components described in the data model. For example, a special equation could describe the loss of a heat pipe. This module must allow applying an equation to a physical component described in the generic data model.
    - Instanced Data Model: It instances the Generic Data Model in order to describe all the physical components (building plans, data centers, servers, air conditioner, etc.) managed by the supervisors connected to RIDER. It is used to store the data representing the real world, which can be used by the data warehouse.
    - ETL: This component will collects and process all (normalised) data coming from various sources, many of it being real time, « stream » data.
    - Data Mining: The data mining component extracts all relevant knowledge from the data warehouse, and provides it to the application services. Descriptive (summary, classification, frequent patterns extraction) and predictive (interpolation, etc.) functionalities must be provided, that match the processing applications' needs.

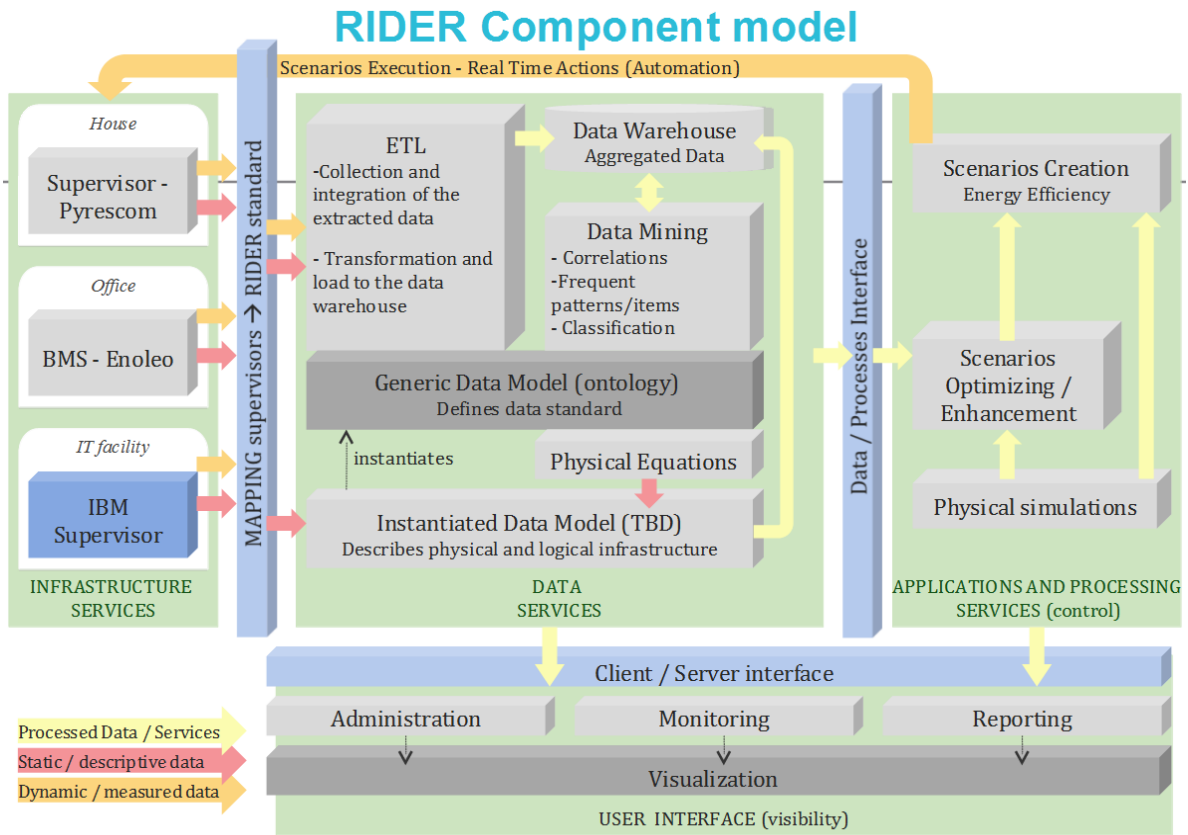

Figure 8. RIDER Component Model

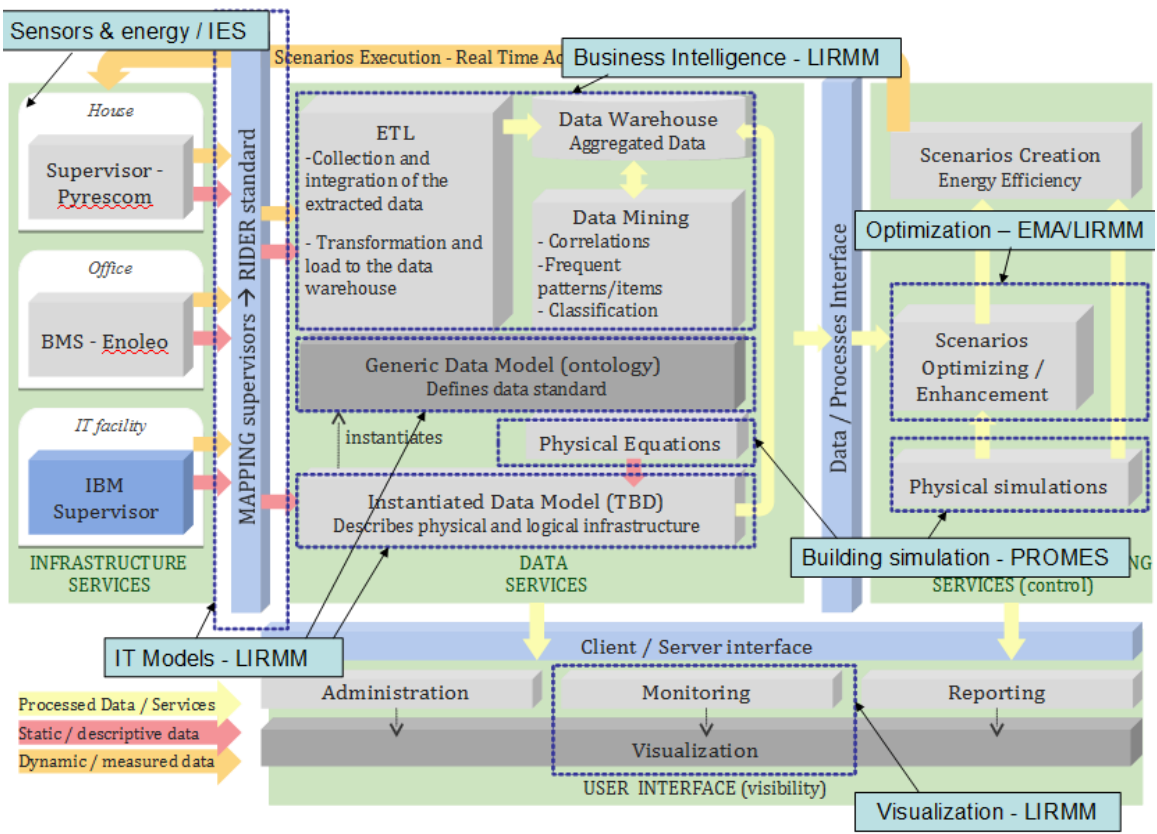

Figure 9. Scientific work-streams positioned

## 5. CONCLUSION

We presented in this paper an approach answering the various requirements of our project context (Research for IT Driven Energy Efficiency). Several scientific work streams were identified: Modelling for multi-scale & multi-standard, very large volumes of data coming from sensors, real time data feeds from high volume of sensors, Business intelligence and decision support, data acquisition and transmission across distances for a small energy cost, energy management – optimisation algorithms and interactive visualisation of heterogeneous data in 3D.